\documentclass[prl,twocolumn,longbibliography,amsmath,superscriptaddress,amssymb]{revtex4-1}

\usepackage{graphicx}
\usepackage{dcolumn}
\usepackage{bm}
\usepackage{amssymb}
\usepackage{amsmath}
\usepackage{color}

\usepackage{placeins}
\usepackage{epstopdf}

\newcommand{\rv} {{\mathbf r}}
\newcommand{\kv} {{\mathbf k}}
\newcommand{\Kv} {{\mathbf K}}
\newcommand{\nv} {{\mathbf n}}

\newcommand{\KP} {${\mathbf k}\!\cdot\!{\mathbf p}$~}

\newcommand{\bra}[1]{\left\langle\,#1\,\right|}
\newcommand{\ket}[1]{|\,#1\,\rangle}
\newcommand{\braket}[2]{\langle\,#1\,|\,#2\,\rangle}

\newcommand{\VEC}[1]{\mathbf{#1}}

\usepackage[normalem]{ulem}

\begin{document}

\title{Spin Filtering via Resonant Reflection of Relativistic Surface States}
\date{\today}

\author{I. A. Nechaev}
 \affiliation{Centro de F\'{i}sica de Materiales CFM -- MPC and Centro Mixto CSIC-UPV/EHU, 20018 San Sebasti\'{a}n/Donostia, Spain}
 \affiliation{Tomsk State University, 634050 Tomsk, Russia}
 \affiliation{Saint Petersburg State University, 198504 Saint Petersburg,  Russia}

\author{E. E. Krasovskii}
\affiliation{Departamento de F\'{i}sica de Materiales, Facultad de Ciencias Qu\'{i}imicas, Universidad del Pais Vasco/Euskal Herriko Unibertsitatea,
Apdo. 1072, San Sebasti\'{a}n/Donostia, 20080 Basque Country, Spain}
\affiliation{Donostia International Physics Center (DIPC), Paseo Manuel de Lardizabal 4, San Sebasti\'{a}n/Donostia, 20018 Basque Country, Spain}
\affiliation{IKERBASQUE, Basque Foundation for Science, 48013 Bilbao, Spain}

\begin{abstract}
A microscopic approach is developed to scattering of surface states from a non-magnetic
linear defect at a surface with strong spin-orbit interaction. Spin-selective reflection
resonances in scattering of Rashba-split surface states by an atomic stripe are theoretically
discovered in a proof-of-principle calculation for a model crystal potential. Spin-filtering
properties of such linear defects are analyzed within an envelope-function formalism for a
perturbed surface based on the Rashba Hamiltonian. The continuous Rashba model is found to
be in full accord with the microscopic theory, which reveals the essential physics behind
the scattering resonance. The spin-dependent reflection suggests a novel mechanism to
manipulate spins on the nanoscale.
\end{abstract}
\maketitle

Scattering of spin-orbit coupled electrons by extended defects arises in many
spintronics-related phenomena, such as spin transport, accumulation, and filtering, which
underlie the manipulation of spin currents in spin-based devices~\cite{Wolf2001,Bercioux2015}.
Furthermore, a detailed understanding of reflection and transmission of relativistic electrons
is important for the unambiguous interpretation of scanning tunneling spectroscopy of
spin-orbit split surface states~\cite{El-Kareh2013,Schirone2015}. Similar problem arises in
ballistic transport through interfaces where powerful {\it ab initio} methods exist for
scattering of bulk electrons from surfaces, such as multiple scattering~\cite{Henk2006},
embedded Green-function~\cite{Wortmann2002}, or Bloch-waves formalism~\cite{Krasovskii2004}.
These methods are, however, not directly applicable to scattering of surface states because
of a complicated structure of the incident and reflected waves in the asymptotic (unperturbed)
region. Therefore, scattering of surface states has been considered either within a
tight-binding scheme \cite{Kobayashi2011} or within a \KP theory combined with continuity
conditions for the envelope function \cite{Usaj2005,Reynoso2006,Sablikov2007,Xie2016} (see
also the application to spin dependent transport in nanowires~\cite{Zhang2005,Alomar2016}).
However, in the \KP method the smoothness of the envelope spinor function generally
conflicts with current conservation~\cite{Molenkamp2001}, which leads
to a qualitatively incorrect separation of the probability current into the
spin-orbit and classical-momentum contributions~\cite{Krasovskii2014}.
The tight-binding formalism, on the other hand, is not well suited for
free-electron-like motion along the surface. This calls for a more
universal approach to scattering of two-dimensional (2D) states, which
could be formulated in an {\it ab initio} framework.

We present a method to microscopically calculate the scattering of spin-orbit split 2D
states from a linear (1D-periodic) defect. ``Microscopic'' means that the system is defined by
the crystal potential $V(\rv)$, and the wave functions satisfy the Schr\"odinger equation in
real space. Therefore, the method can be straightforwardly transferred to {\it ab initio}
calculations. Here, we report a proof-of-principle calculation of the transmission of
Rashba-split states through atomically thin defects. We study spin-filtering properties of the
defects and discover spin-dependent reflection resonances for certain scatterers.

Previous studies of the effect of spin-orbit coupling on the scattering of 2D states included
multibeam spin-polarized reflection from a lateral
barrier~\cite{Govorov2004, Chen2005, Teodorescu2009},
spin accumulation at the edges of semi-infinite
systems~\cite{Usaj2005, Reynoso2006, Sonin2010, Khaetskii2013},
spin selective refraction at an interface of two 2D media~\cite{Khodas2004}, spin dependent
transmission of electrons incident from a non-relativistic medium through a barrier with
spin-orbit coupling~\cite{Ramaglia2004, Sablikov2007}, and a semi-classical reflection
from a smooth barrier~\cite{Silvestrov2006}. The above studies relied on an
envelope-function description of the surface states using effective Hamiltonians.
By contrast, here, the perturbed surface is treated fully microscopically: the scattering
problem is reduced to a supercell band structure problem, which naturally involves both the
propagating and all the required evanescent 2D waves and yields a
detailed description of scattering, beyond the envelope-function
picture. Still, the resonant properties of the scatterer can be related to the
parameters of a \KP Rashba model for the perturbed surface. This demonstrates the
generality of the phenomenon and suggests a way to its experimental realization.

\begin{figure*}[t]
\includegraphics[trim =0mm 0mm 0mm 0mm, clip=true, width=\textwidth]{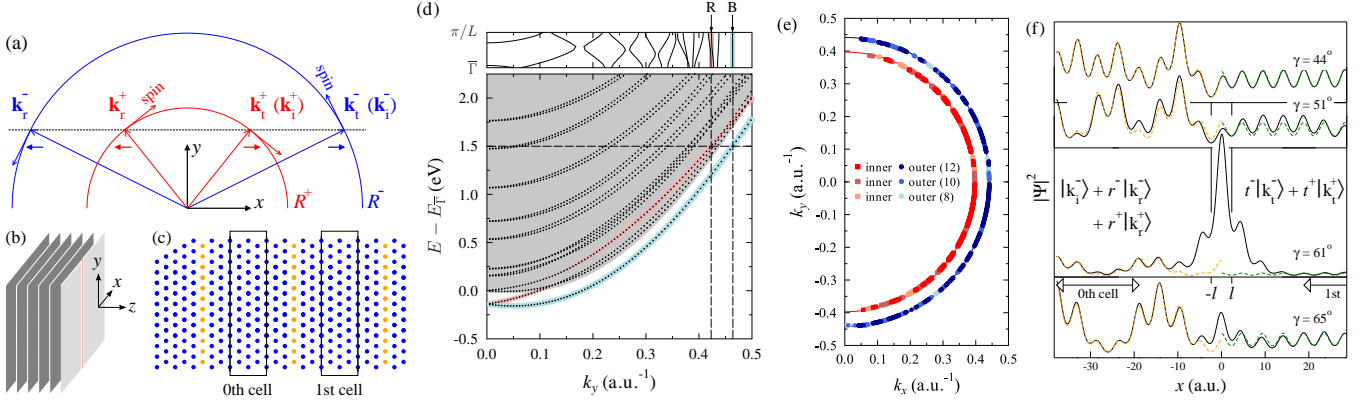}
\caption{\label{FIG1}
  (a) CECs of Rashba split surface states. $\kv^\pm_{\rm r}$, $\kv^\pm_{\rm t}$, and
    $\kv^\pm_{\rm i}$ are the Bloch vectors of reflected, transmitted, and incident waves.
(b) Finite-thickness slab with a linear defect in the topmost layer.
(c) Supercell geometry: topmost layer with a repeated row of impurity atoms. The
    two boxes indicate two asymptotic regions.
(d) Upper panel: CEC at $E-E_{\bar\Gamma}=1.5$~eV for 12-fold supercell. Lower panel:
    dots are the dispersion $E(K_y)$ of the solutions $\Phi_{\Kv^n}$ for $K_x=0$.
    Shaded area shows the $k_y$-projected states of the ideal surface. The
    spin-orbit split states due to the defect are highlighted blue (true bound state)
    and red (resonance when inside the gray area). In the 1.5~eV
    CEC, arrows indicate the bound state (B) and the resonance (R).
(e) Bloch vectors $k_x$ extracted from the eigenvalues $\exp(i\boldsymbol\tau\kv)$ of the
    host lattice translation operator for the three supercells for the well ($\epsilon=1.04$, upper quadrant) and the barrier
    ($\epsilon=0.96$, lower quadrant).
    (f) Density profiles of four scattering states at $E-E_{\bar\Gamma}=0.7$~eV for
    $\gamma=44$, 51, 61, and 65$^\circ$ for the well $\epsilon=1.07$ in a 12-fold supercell.
    Dashed lines are their asymptotic representations in 0th and 1st cells continued up to the defect.
}
\end{figure*}
Typical constant energy contours (CEC) of Rashba-split states comprise two circles centered at
$\bar\Gamma$ with spin oriented along $\kv\times\nv$ for the inner circle (of radius
$R^+$) and along $-\kv\times\nv$ for the outer circle ($R^-$), where $\nv$ is the surface
normal and $\kv$ is the 2D Bloch vector. We denote the unperturbed states by
$\ket{\kv^\chi_\xi}$, where $\chi=\pm$
indicates chirality and $\xi=\rm r/t$ is the
propagation direction along $x$, see Fig~\ref{FIG1}(a). Consider a defect created by
substituting a row of atoms (along $y$ axis) by a different atom, Fig.~\ref{FIG1}(b).
For a surface state $\ket{\kv_{\rm i}^\pm}$ incident from the left half-plane the scattering
solution $\ket{\Psi}$ far from the defect contains two transmitted $\ket{\kv^\pm_{\rm t}}$
and two reflected $\ket{\kv^\pm_{\rm r}}$ waves, Fig.~\ref{FIG1}(a).
The crystal momentum along $y$ is conserved, so
$k^+_{{\rm t}y}=k^-_{{\rm t}y}=k^+_{{\rm r}y}=k^-_{{\rm r}y}=k^\chi_{{\rm i}y}$.
For $R^->k^-_{{\rm i}y}>R^+$ there is only one transmitted and one reflected wave. In the
unperturbed region, $\ket{\Psi}$ contains also evanescent waves, and depending on how fast they
decay away from the defect the scattering state $\ket{\Psi}$ can be
obtained from band structure solutions $\Phi_{\Kv^n}$
for a smaller or larger supercell, Fig.~\ref{FIG1}(c). The method works as follows: At a given
$k_{{\rm i}y}^\chi$, the surface states perturbed by the periodic defect give rise to four (or two)
supercell eigenfunctions $\Phi_{\Kv^n}$ with the supercell crystal momentum
$K_y^n=k^\chi_{{\rm i}y}$, $n=1,2,3,4$ ($n=1,2$ if $k^{-}_{{\rm i}y}>R^+$), see Fig.~\ref{FIG1}(d).
Far from the scatterers [in the asymptotic region denoted 0th cell in Fig.~\ref{FIG1}(c)]
the functions $\Phi_{\Kv^n}(\rv)$ can obviously be decomposed into a sum of the four
unperturbed surface states $\ket{\kv_\xi^\chi}$.
The latter are obtained as eigenfunctions of the translation
operator of the ideal surface
$\hat T\Theta(\rv)=\Theta(\rv+\boldsymbol\tau)=\exp(i\boldsymbol\tau\kv)\Theta(\rv)$
in terms of the supercell solutions: $\Theta=\sum_n c^n\Phi_{\Kv^n}$. The functions $\Theta$
are defined everywhere in the crystal, and in the asymptotic region they
coincide with the unperturbed surface states: $\Theta_\xi^{\chi}(\rv)=\braket{\rv}{\kv_\xi^\chi}$.
The full scattering solution is then a linear combination
$\Psi=\sum_{\chi\xi} a^\chi_\xi\Theta_\xi^\chi$ defined by the condition that $\Psi$ contain only
one right-traveling wave in the 0th supercell and no left-traveling waves in the 1st supercell
(the next asymptotic region), Fig.~\ref{FIG1}(c). Note that $\Psi$ is valid everywhere, including the vicinity of the defect.

\begin{figure}[t]
\includegraphics[width=0.85\columnwidth]{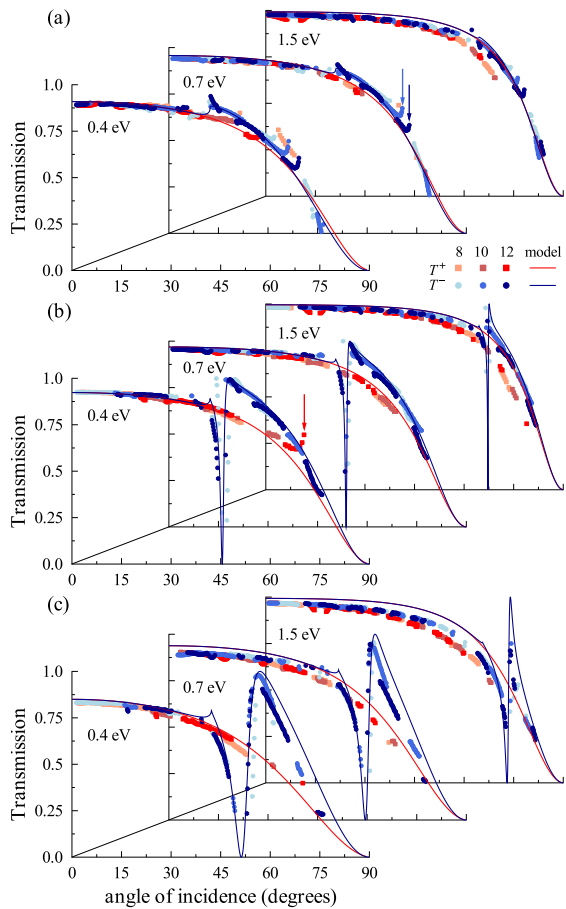}
\caption{\label{FIG2} Transmission probability as a function of the angle of incidence for
the $\epsilon=0.96$~(a), 1.04~(b), and 1.07~(c) for $E-E_{\overline{\Gamma}}=1.5$, 0.7, and
0.4~eV. The shades of red (blue) show $T^+$ ($T^-$) for 8, 10, and 12-fold supercells. Solid
lines are the continuous model fit of $T^+$ (red) and $T^-$ (blue) obtained with $U=0.27$,
$-0.33$, and $-0.55$~eV for $\epsilon=0.96$, 1.04, and 1.07, respectively. The parameters
$\alpha$ and $m^*$ for the presented energies are listed in Table~\ref{tab:model_params}.
}
\end{figure}

Let us consider a 7-layer slab with the geometry of a Au(111) surface with an overlayer.
The atoms are represented by a 3D regular muffin-tin potential, which is expanded in a
truncated 3D Fourier series and included into the microscopic Hamiltonian
${\hat p}^2 + V(\VEC{r}) + \beta{\boldsymbol\sigma}\cdot
\left [\,\boldsymbol\nabla V({\VEC r})\times\VEC{\hat p}\,\right ]$,
with $\beta$ scaled such that the Rashba splitting of the surface states be close to
that in Au(111). The supercell band structure is calculated on a rectangular $\kv$-mesh with
$\Delta K_x=\Delta K_y=0.0056$~a.u.$^{-1}$, and the functions $\Phi_{\Kv^n}$ for a given energy
and $K_y$ are obtained by triangular interpolation. A typical constant energy contour is shown
in the upper panel of Fig.~\ref{FIG1}(d). The artificial periodicity of the defect
gives rise to spectral gaps, so for certain $K_y$ there are no solutions $\Phi$. However, for
a given $K_y$ one can always choose a supercell for which the solution exists.

We will consider two types of defects: barrier and well. For a barrier, the potential at the
impurity site $U_{\rm D}$ is shallower than the potential $U_{\rm S}$ at the host atom, and for
a well it is deeper. Computationally, the muffin-tin potentials $U_{\rm D}$ are linearly scaled:
$U_{\rm D}(r)=\epsilon U_{\rm S}(r)$. The Bloch vectors of the unperturbed surface states
extracted from the eigenvalues $\exp(i\boldsymbol\tau\kv)$ are shown in Fig.~\ref{FIG1}(e).
The good agreement between the three supercells both for a well and for a barrier demonstrates
that the evanescent waves are negligible in the asymptotic region already for the 8-fold
supercell. Figure~\ref{FIG1}(f) shows the density profiles of the outer-circle surface states
at $E-E_{\bar\Gamma}=0.7$~eV scattered by a well $\epsilon=1.07$ for four angles of incidence
$\gamma$. Although the unperturbed surface
states are derived from 0th cell, the asymptotic representation is seen to be valid over a
much wider region (see, especially, $\gamma=44^\circ$). The transmission
probability $T^\pm$ as a
function of $\gamma$ is shown in Fig.~\ref{FIG2} for a barrier, $\epsilon=0.96$, and for two
wells, $\epsilon=1.04$ and 1.07. Here $T^+$ and $T^-$ stand for the incident wave in the inner
and in the outer circle, respectively. The colored symbols are the microscopic calculations,
with shades of red used for $T^+$ and blue for $T^-$. The voids
in the curves correspond to the gaps in the supercell band structure, and in approaching the gap the
transmission sometimes shows a spurious growth [see vertical arrows in Figs.~\ref{FIG2}(a) and
\ref{FIG2}(b)]. This happens when two of the Bloch vectors $K_y^n$ are close to the edge of the
Brillouin zone, and the numerical method finds the two solutions $\Phi_{\Kv^n}$ linearly
dependent. Such artifacts are recognized by an accuracy criterion, and they are easily sorted
out because they occur at different angles for different supercells.

\begin{table}
\caption{\label{tab:model_params}
Rashba Hamiltonian parameters $m^{\ast}$ and $\alpha$ used to model the transmission
through the defect, Fig~\ref{FIG2}. They are derived
by fitting the dispersion $E(k)$ of the unperturbed surface state of the microscopic
calculation. $m^{\ast}$ and $\alpha$ depend on energy because $E(k)$ is not exactly
parabolic. Atomic Hartree units are used: $\hbar=m_0=e=1$.}
\begin{ruledtabular}
\begin{tabular}{cccc}
$E-E_{\overline{\Gamma}}$ (eV) & $m^{\ast}$ (a.u.) & $\alpha$ (a.u.) & $R^--R^+$ (a.u.$^{-1}$) \\
\hline
   0.0 &  1.59 &               0.033 &       0.105 \\
   0.4 &  1.85 &               0.026 &       0.096 \\
   0.7 &  2.00 &               0.020 &       0.080 \\
   1.5 &  2.00 &               0.012 &       0.048 \\
\end{tabular}
\end{ruledtabular}
\end{table}

Most important is the strikingly different behavior of the transmission probability $T^-$ for
the two types of defects: for a barrier, $T^-$ steadily decreases, Fig.~\ref{FIG2}(a), whereas
for a well it shows a sharp minimum followed by a maximum, see Figs.~\ref{FIG2}(b) and
\ref{FIG2}(c). By contrast, $T^+$ steadily decreases in both cases. To understand this
behavior, let us consider the contribution of evanescent waves to the
scattering states $\Psi$. Their weight can be inferred from the deviation of the density
profile $|\Psi|^2$ in Fig.~\ref{FIG1}(f) from the left and right asymptotics (dashed lines)
continued up to the scatterer. For small angles the weight of the evanescent waves is
negligible, and it starts growing when $k^{-}_{{\rm i}y}$ exceeds $R^+$ because the evanescent
waves replace the missing propagating solutions of the inner circle. This point manifests
itself by a cusp maximum in $T^-$, e.g., at 0.4~eV around $\gamma=45^\circ$ in
Fig.~\ref{FIG2}(a). In approaching the minimum [see $\gamma=61^\circ$ in Fig.~\ref{FIG1}(f)]
the density around the defect steeply grows and then rapidly decreases with increasing
$\gamma$. This happens because the defect causes a sharp perturbation of the potential $V(\rv)$
(comparable to the lattice period), which is known to give rise to a bound state localized at
the defect and energetically split off from the band continuum \cite{Madelung1978}. For
the Rashba states that are bounded only from below a barrier does not produce any bound states.
By contrast, a well-like perturbation produces two structures [Fig.~\ref{FIG1}(d)]: bound
state B and its spin-orbit counterpart resonance R [highlighted red in Fig.~\ref{FIG1}(d)].
The hybridization of the incident wave of the outer circle with the
resonance -- the inner branch of the Rashba split 1D
impurity state -- gives rise to the asymmetric $T^{-}(\gamma)$ feature.

In order to relate the reflection resonance to phenomenologically relevant spin-orbit
characteristics of the material let us consider a Rashba system with the Hamiltonian
$\hat H_{\mathrm{R}}=k^2/2m^{\ast}+\alpha(k_y\sigma_x-k_x\sigma_y)$. The relation $k^2=k_x^2+k_y^2$
determines whether a given branch is propagating or evanescent for a given $k_y$ and
$E$~\cite{Sablikov2007}. The defect is represented by a potential barrier (well) $V(x)=U$
for $-l<x<l$, with $V(x)=0$ elsewhere. The width of the defect equals the width of the unit
cell: $2l=\tau_x$. The scattering solution is found by the condition of the continuity of the
spinor wave function and flux across the defect~\cite{Molenkamp2001}. Thus, the
envelope-function formalism solves the problem without resorting
to an artificial supercell. A four-wave representation in the perturbed region
$\sum_{\chi\xi}d^{\chi}_{\xi}\ket{\tilde{\mathbf{k}}^{\chi}_{\xi}}$ is matched to the wave function
in the left and right half-planes at the boundaries $x=\pm l$ indicated in Fig~\ref{FIG1}(f).
Here $\tilde{\mathbf{k}}$ are the wave vectors of the eigenfunctions of the Hamiltonian
$\hat H_{\mathrm{R}}+U$ (with the same $\alpha$ and $m^{\ast}$ as
for the unperturbed surface). The scattering problem then reduces to an $8\times 8$
matrix equation $\hat{M}\mathbf{a}=\mathbf{f}$ for the vector
$\mathbf{a}=(r^{\pm}, d^{\pm}_{\mathrm{r}}, d^{\pm}_{\mathrm{t}}, t^{\pm})^{\mathrm{T}}$. Here
$\hat{M}$ is the matching matrix, and $r^{\pm}$ and $t^{\pm}$ are the coefficients of the
two reflected and two transmitted waves in the two unperturbed half-planes, see the
legends in Fig.~\ref{FIG1}(f). The right-hand side $\mathbf{f}$ represents the
incident wave $\ket{\kv_{\rm i}^\chi}$, and it has four non-zero components: the value and
the flux (for both spins) at $x=-l$.
\begin{figure}[t]
\includegraphics[width=0.93\columnwidth]{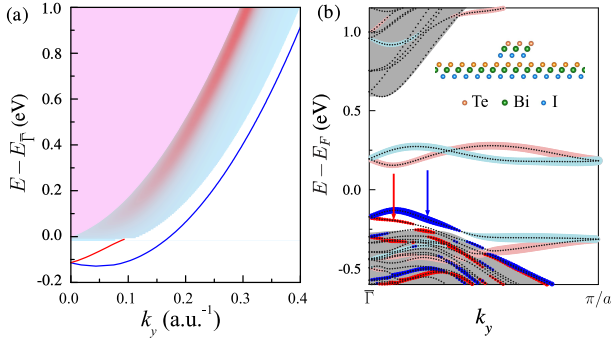}
\caption{\label{FIG3} (a) Electronic structure of the Rashba system with linear
defect. Solid lines show the bound states split off from the Rashba continuum.
The width of the blurred red line shows the $k_y$-width of the resonance. The
Rashba continuum of the ideal system $E^{\pm}({\mathbf{k})}$ [cf. grey area in
Fig.~\ref{FIG1}(d)] is shown by the sign of the $k_y$-projected $x$-spin spectral
density $S_x^{\rm tot}$. Light red indicates $S_x^{\rm tot}>0$, and light blue
$S_x^{\rm tot}<0$.
(b) Band structure and side view geometry of a BiTeI trilayer with a nanostripe. Shaded area
covers the $k_y$-projected states of the clean trilayer. The localization of the trilayer
states on the atoms  beneath the stripe are shown by red ($\sigma_x^{\uparrow}$) and
blue ($\sigma_x^{\downarrow}$) fat bands. Light red (light blue) fat bands correspond to
$\sigma_x^{\uparrow}$ ($\sigma_x^{\downarrow}$) dangling bond states localized on the stripe.
}
\end{figure}

The transmitted current $T^\chi=|t^{+}_\chi|^2 + |t^{-}_\chi|^2$ is shown in Fig.~\ref{FIG2}
by solid lines. With $U$ adjusted to fit the microscopic calculations the Rashba model
perfectly reproduces the shape of the curves and the dependence of the position
and the width of the resonance on the energy and on the scatterer. Surprisingly,
the envelope-function method originally designed for slowly varying potentials
shows excellent performance for the atomic stripe. To establish the analogy with
the microscopic picture, let us consider the eigenspectrum of the perturbed system. It is
obtained by dropping the incident wave and finding zeros of real and imaginary part of the
determinant of the matrix $\hat{M}$. In Fig.~\ref{FIG3}(a), the ideal surface is presented
by the energy-momentum distribution of the sign ($\uparrow$ or $\downarrow$)
of the $k_y$-projected $\sigma_x$-spin spectral density $S_x^{\rm tot}=S^{+}_x+S^{-}_x$, where
$S^{\pm}_x(E, k_y)=\int dk_x \bra{\mathbf{k}^{\pm}} \sigma_x\ket{\mathbf{k}^{\pm}}
\,\delta[E-E^{\pm}({\mathbf{k})}]/8\pi^2$.
The incident wave comes from the $\sigma_x^{\downarrow}$ continuum (blue area),
which overlaps with the spectral resonance having $\sigma_x^{\uparrow}$ spin. Just at the
resonance, $T^-(\gamma)$ sharply drops to zero and then steeply rises to unity, exactly
as in the microscopic model, see Figs.~\ref{FIG2}(b) and \ref{FIG2}(c). Thus,
the scattering by the 1D defect is transparently expressed through the relation
between the CECs of the host and the defect region. Here, an important ingredient
is the spin non-conservation, so the effect does not occur, e.g., for Zeeman splitting.

\begin{figure}[t]
\includegraphics[width=0.97\columnwidth]{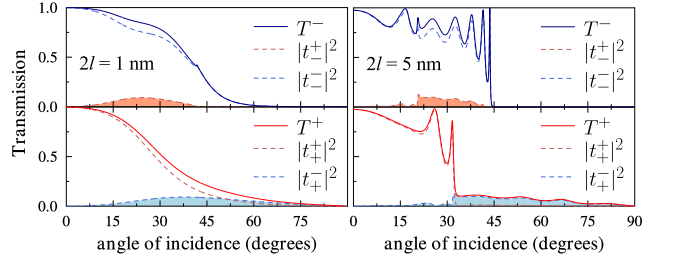}
\caption{\label{FIG4} Transmission through barriers of width $2l=1$~nm and 5~nm with
$U=0.27$~eV. Dashed lines show the partial currents $|t^{+}_\chi|^2$ (red) and
$|t^{-}_\chi|^2$ (blue). The current carried by the wave of the opposite chirality
to the incident wave is shown by the shaded areas.
}
\end{figure}

For the above monoatomic stripes, practically all the transmitted current is carried
by the same wave as is incident, so the spin orientation of the incident electron is
preserved on transmission. The picture becomes
very different for thicker stripes, in which the evanescent waves (complex $\tilde{k}_x$) do
not participate in the transmission through the defect. The continuous Rashba model predicts
that for barriers thicker than 1~nm the spin-flip transmission, i.e., $\ket{\kv^\mp_{\rm t}}$
for $\ket{\kv^\pm_{\rm i}}$, becomes quite important, see Fig.~\ref{FIG4}. Another vivid
feature of the nanosized barrier are the Fabry-P\'{e}rot oscillations of the transmission.

Finally, as a possible platform for the experimental realization of the discovered
resonant reflection, we suggest the layered semiconductors of the BiTe$X$
($X=$I, Br, Cl) family, where giant Rashba-split surface states reside in an absolute
gap~\cite{Ishizaka2011, Crepaldi2012, Eremeev2012, Sakano2013, Eremeev2013}. Already a
trilayer Te-Bi-$X$ -- the easily exfoliated structure element of these
semiconductors -- provides the desired 2D spin-orbit split valence and conduction
states~\cite{Chen2013,Ma2014}. For holes and electrons of a stand-alone
trilayer, a perturbation can be introduced by putting on it a nanostripe as
shown in the inset of Fig.~\ref{FIG3}(b)~\footnote{We consider a stripe of one-nanometer
width made of a single BiTeI trilayer and with stable stoichiometric edges
as in Ref.~\cite{Eremeev2017}.}. This linear defect gives rise to 1D spin-orbit split bound
states that split off from the valence band, as follows from our {\it ab initio}
calculation~\footnote{Our DFT-GGA calculations employed the full-potential
linearized augmented-plane-wave method implemented in the FLEUR code,
http://www.flapw.de.}, see vertical arrows in Fig.~\ref{FIG3}(b). These states form the 1D Rashba channel that guides
the holes. As seen in Fig.~\ref{FIG3}(b), the inner branch (red arrow) becomes the
above-mentioned spectral resonance when it enters the projected continuum.

To summarize, we have developed a microscopic approach to scattering
of relativistic surface states by a linear defect and found strong spin
selectivity of electron transmission for well-like perturbations.  Thereby, the
transmitted spin current can be enhanced, which suggests a potential
technique for non-magnetic spin filtering and spin injection.

\begin{acknowledgments}
This work was supported by the Spanish Ministry of Economy and Competitiveness MINECO
(Project No. FIS2016-76617-P). I.A.N. also acknowledges support from Saint Petersburg
State University (Grant No. 15.61.202.2015).
\end{acknowledgments}

\end{document}